\documentclass[aps,prl,twocolumn,superscriptaddress]{revtex4}
\usepackage{graphicx,subfigure}
\usepackage{hyperref}
\usepackage{amsmath,amssymb}
\usepackage{multirow}
\usepackage{textcomp}
\usepackage{float}
\usepackage{color}

\newcommand{\ket}[1]{\ensuremath{\vert #1 \rangle}}

\newcommand{\abs}[1]{\ensuremath{\left\vert #1 \right\vert}}

\renewcommand{\vec}[1]{\ensuremath{\boldsymbol{\mathbf{#1}}}}
\newcommand{\uvec}[1]{\ensuremath{\hat{\mathbf{#1}}}}

\renewcommand{\imath}{\ensuremath{\mathrm{i}}}
\newcommand{\phis}{\ensuremath{\alpha}}

\newcommand{\Vmod}{\ensuremath{\tilde{V}}}

\usepackage{ulem}

\newcommand{\editout}[1]{{\bf \sout{#1}}}      

\renewcommand{\editout}[1]{}

\begin{document}
\title{Dynamic Optical Superlattices with Topological Bands}
\author{Stefan K. Baur}
\affiliation{T.C.M. Group, Cavendish Laboratory, J.J. Thomson Avenue, Cambridge CB3 0HE, United Kingdom}
\author{Monika H. Schleier-Smith}
\affiliation{Department of Physics, Stanford University, Stanford, California 94305, USA}
\affiliation{Fakult\"{a}t f\"{u}r Physik, Ludwig-Maximilians-Universit\"{a}t, Schellingstrasse 4, 80799 M\"{u}nchen, Germany}
\affiliation{Max-Planck-Institut f\"{u}r Quantenoptik, Hans-Kopfermann-Str. 1, 85748 Garching, Germany}
\author{Nigel R. Cooper}
\affiliation{Cavendish Laboratory, J.J. Thomson Avenue, Cambridge CB3 0HE, United Kingdom}
\affiliation{Fakult\"{a}t f\"{u}r Physik, Ludwig-Maximilians-Universit\"{a}t, Schellingstrasse 4, 80799 M\"{u}nchen, Germany}

\begin{abstract}
We introduce an all-optical approach to producing high-flux synthetic magnetic fields for neutral atoms or molecules by designing intrinsically time-periodic optical superlattices. A single laser source, modulated to generate two frequencies, suffices to create dynamic interference patterns which have topological Floquet energy bands. We propose a simple laser setup that realizes a tight-binding model with uniform flux and well-separated Chern bands.  Our method relies only on the particles' scalar polarizability and far detuned light.
\end{abstract}
\date{\today}

\maketitle

In the quest to establish ultracold atoms as versatile quantum simulators of condensed-matter physics \cite{Bloch:2012uq}, a key challenge is to develop minimally invasive methods of mimicking the orbital effects of a magnetic field \cite{dalibardreview}.  In solid-state systems the interplay of strong magnetic fields with Coulomb interactions gives rise to strongly correlated phases, notably the fractional quantum Hall effect \cite{Laughlin:1983}.  Atomic systems offer prospects for studying related phenomena of either bosons or fermions with tunable interactions, using new diagnostic tools.

This goal has motivated intense theoretical and experimental effort at simulating, for neutral atoms, the Lorentz force experienced by a charged particle in a magnetic field \cite{dalibardreview}.  Methods demonstrated to date have included subjecting quantum gases to rapid rotation \cite{Cooper:2008adv,Fetter:2009} or imprinting geometric phases via spatially dependent couplings between internal states \cite{Lin:2009qf}.  While the latter approach can in principle be extended to reach a high (net positive) flux density \cite{Jaksch:2003vn,Mueller:2004ys,GerbierDalibard,Cooper:2011kl,Cooper:2011cr}, only a select few atomic species offer a route to introducing the requisite optical couplings without significant spontaneous-emission-induced heating or atom loss \cite{GerbierDalibard,Cooper:2011kl,Cooper:2011cr,Wei:2013kx,Cui:2013}.

A more broadly applicable means of producing high-flux gauge fields is by modulating tight-binding optical lattices periodically in time \cite{Eckardt:2005tg,Sias:2008kx,Kitagawa10,Kolovsky:2011zr,Hauke:2012fk,Creffield:2013zr,Rudner:2013fv,AidelsburgerKennedy,circshaking}.    This approach can be implemented with arbitrarily far-detuned light, in principle for any atomic or molecular species.  Analogous methods \cite{Oka09,Lindner10,Iadecola2014} have even been applied in solid-state systems~\cite{Wang13} and photonic crystals \cite{Rechtsman13}.  Common to all these systems is a breaking of time-reversal symmetry that endows the tunneling matrix elements with Peierls-like phases \cite{Eckardt:2005tg,Sias:2008kx,Hauke:2012fk,Struck1213}, mimicking the effect of a magnetic flux though each lattice plaquette.

In optical lattices, Peierls phases have been engineered by large-amplitude off-resonant shaking \cite{Sias:2008kx,Struck1213}; or by direct modulation of on-site energies to produce a {\it resonant} photon-assisted hopping between orbitals of distinct lattice sites \cite{Aidelsburger11,AidelsburgerMiyake}.    The latter method, requiring only small modulation amplitude, is less susceptible to (multiphoton) heating processes.  To date, demonstrations of this method in periodic optical lattices lead to zero average flux.  Recent success in producing large uniform flux by resonant photon-assisted hopping \cite{AidelsburgerMiyake} is a technical feat, requiring not only multiple optical lattice lasers but also a strong magnetic field gradient \cite{footnote1}.

Here, we present an all-optical scheme for realizing a tight-binding Hamiltonian with uniform flux.  Underlying our proposal is a new approach to breaking time-reversal symmetry in far-detuned optical lattices, relying only on interference of light at two frequencies readily derived from a single laser. We introduce this approach with a minimalist scheme yielding a Haldane-like model on a honeycomb lattice~\cite{Haldane} before proceeding to our principal proposal, which achieves uniform flux on a triangular lattice.  We show that both schemes yield well-separated topological Floquet bands.

To realize each topological model, we design a dynamic interference pattern constituting a two-dimensional (2D) lattice that evolves periodically in time.  First, using only a single frequency $\omega$ of light, we engineer a static lattice $V(\mathbf{r})$ whose unit cell comprises two sites (A and B) offset by an energy $\hbar \delta$.  Whereas this energy offset suppresses tunneling between A and B, we reestablish the tunneling---with modified phases---by interfering the lattice beams with one additional laser field, generated from the same source at a nearby frequency $\omega+\Omega$ and propagating normal to the 2D plane, to form a modulating lattice $\tilde{V}$ that oscillates at frequency $\Omega$. The combined effect of $V+\tilde{V}$, forms a \textit{dynamic superlattice} which, for $\Omega\approx\delta$, induces chiral hopping. Since the beams at frequency $\omega$ contribute to both the static and the dynamic lattice, these are locked in register.

\begin{figure}
\includegraphics[width=0.9\columnwidth]{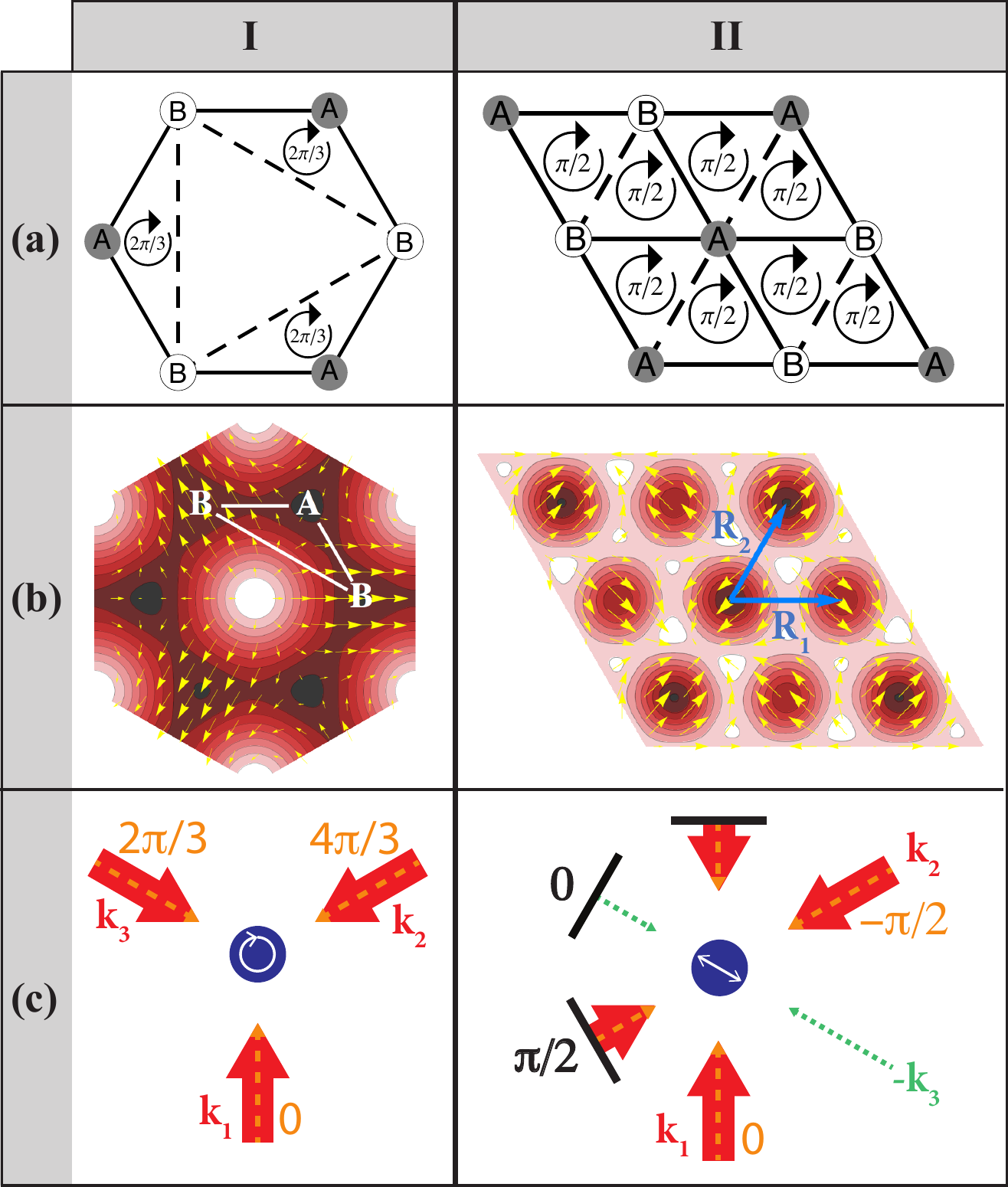}
\caption{\label{fig:overview}(Color online.) Two topological tight-binding models (a) and their realization in dynamic optical superlattices (b-c): (I) Haldane-like model on a honeycomb lattice, formed by three running-wave beams and a circularly polarized coupling laser.  (II) Triangular lattice with uniform flux, formed by two (or three) retroreflected lattice beams and a linearly polarized coupling laser.  In (b), shading indicates the static potential $V$, while arrows indicate the magnitude and phase of the dynamic modulation $\Vmod$.  In (c), thick red (dashed orange) arrows represent the lattice beams' $p$ ($s$) polarization components; orange labels indicate the retardances $\alpha_i$.  In II(c), dotted green lines indicate an optional third standing wave, offset in frequency to avoid interference with the first two, that improves the isotropy of the triangular lattice.  Bold black phases $\varphi_{-2,+3}$ must be stabilized relative to $\varphi_{\pm1,+2,-3}=0$.}
\end{figure}

The two models considered in this work are illustrated in Fig.~\ref{fig:overview}(a), with A/B lattice sites shown as shaded/white circles. Tunneling from A to A or B to B (dashed lines) occurs even within the static lattice, but tunneling from A to B (solid lines) is photon-assisted by the periodic modulations of the dynamic lattice. The full time-dependent potentials are illustrated in Fig.~\ref{fig:overview}(b), where the shading indicates the depth of the static lattice and arrows indicate the amplitude and phase of the modulating superlattice.

To engineer the structure and dynamics of a superlattice produced from a single laser source, we control both polarization components of each of $N$ laser fields
\begin{equation}\label{eq:Ei}
\vec{E}_i = E\left(\cos\gamma\; e^{\imath\phis_i}\uvec{z}+\sin\gamma\; \uvec{k}_i\times\uvec{z}\right)e^{\imath(\vec{k}_i\cdot\vec{r}-\omega t-\varphi_i)}
\end{equation}
forming the static lattice, as illustrated in Fig. \ref{fig:overview}(c).  Here, each phase $\alpha_i$, labeled in orange (gray), denotes a retardance between in-plane ($p$) and out-of-plane ($s$) fields. The parameter $\gamma$ sets the ratio between the amplitudes of the $p$ and $s$ components of $\mathbf{E}_i$. Whereas the $s$ fields contribute only to the static lattice, the $p$ components additionally interfere with a coupling field $\mathbf{E}= \vec{E}_0 e^{i[kz- (\omega+\Omega)t]}$ to produce the modulating lattice $\Vmod$.  For $N\le 3$, as in our honeycomb lattice (I), the resulting dynamic superlattice is insensitive to the relative phases $\varphi_i$ of different lattice beams, which have no effect but to translate the entire structure. Our triangular lattice (II), formed from $N>3$ wavevectors, offers greater control over the modulation pattern via additional phases $\varphi_i$, which can be stabilized as in Ref. \cite{Wirth:2011uq}.  In Fig. \ref{fig:overview}.II.c, we indicate in bold these phases, chosen to produce a flux of $\pi/2$ through each plaquette.

Achieving non-zero fluxes through superlattice plaquettes requires designing a modulation potential $\Vmod$ that breaks the symmetry between clockwise and counterclockwise hopping.  In particular, the on-site modulation $\tilde{V}(\mathbf{R},t)=\tilde{V}_{\mathbf{R}} \cos(\phi_{\mathbf{R}}+\Omega t)$ of the potential at lattice sites $\vec{R}$, with frequency $\Omega \approx \delta$, will: (i) restore photon-assisted tunneling between A and B sites (which are offset in energy by $\hbar \delta$) providing a Peierls
-like tunneling phase; and (ii) modify the amplitude of tunneling between degenerate sites (e.g. from an A site to a neighboring A site).  While the Peierls-like phases determine the magnetic flux, the tunneling amplitudes set the energy scales for the band structure of each dynamic superlattice.

A pictorial method of deriving the photon-assisted tunneling matrix elements is to draw the vectors corresponding to the complex numbers $z_{\mathbf{R}}=\tilde{V}_{\mathbf{R}} e^{i \phi_{\mathbf{R}}}/\Omega$. Tunneling from a higher energy (B) site at $\mathbf{R}$ to lower energy (A) site at $\mathbf{R}'$ is described by the difference vector $z=z_{\mathbf{R}}-z_{\rm \mathbf{R}'}$: the phase $\theta$ and amplitude ${\cal A}>0$ of $z\equiv{\cal A} e^{i \theta}$ determine the effective tunneling matrix element \cite{Kolovsky:2011zr, Hauke:2012fk}
\begin{eqnarray}
K_{\rm eff}= 
 e^{i \theta} J_1({\cal A}) \times K \,.
\label{eq:keff}
\end{eqnarray}
Here, $K$ is the bare tunneling matrix element for a time-independent Hamiltonian in which the two sites are degenerate, and we have assumed $K\ll\Omega,\delta$. While the phases $\theta$ given by Eq. \ref{eq:keff} are only defined up to a gauge transformation, the magnetic flux through a plaquette, given by the sum of the Peierls-like phases around that plaquette (divided by $2\pi$), is gauge invariant. For degenerate lattice sites (e.g. tunneling from A to A), the tunneling matrix element $J\ll\Omega,\delta$ for the static lattice is renormalized to
\begin{eqnarray}
J_{\rm eff}= J_0({\cal A}) \times J\,.
\label{eq:jeff}
\end{eqnarray}
Here, $J_{0,1}(x)$ denote Bessel functions. In the limit of small amplitude of the resonant modulation, ${\cal A} \lesssim 1$, in which case $K_{\rm eff}\simeq K {\cal A}e^{i\theta}/2$ and $J_{\rm eff}\simeq J$.
Using the rules (\ref{eq:keff}) and (\ref{eq:jeff}), we derive the effective time-independent tight-binding Hamiltonian for each of the schemes in Fig. \ref{fig:overview}(c), before proceeding to a full calculation of the topological band structure.  
\begin{figure}
\includegraphics[width=\columnwidth]{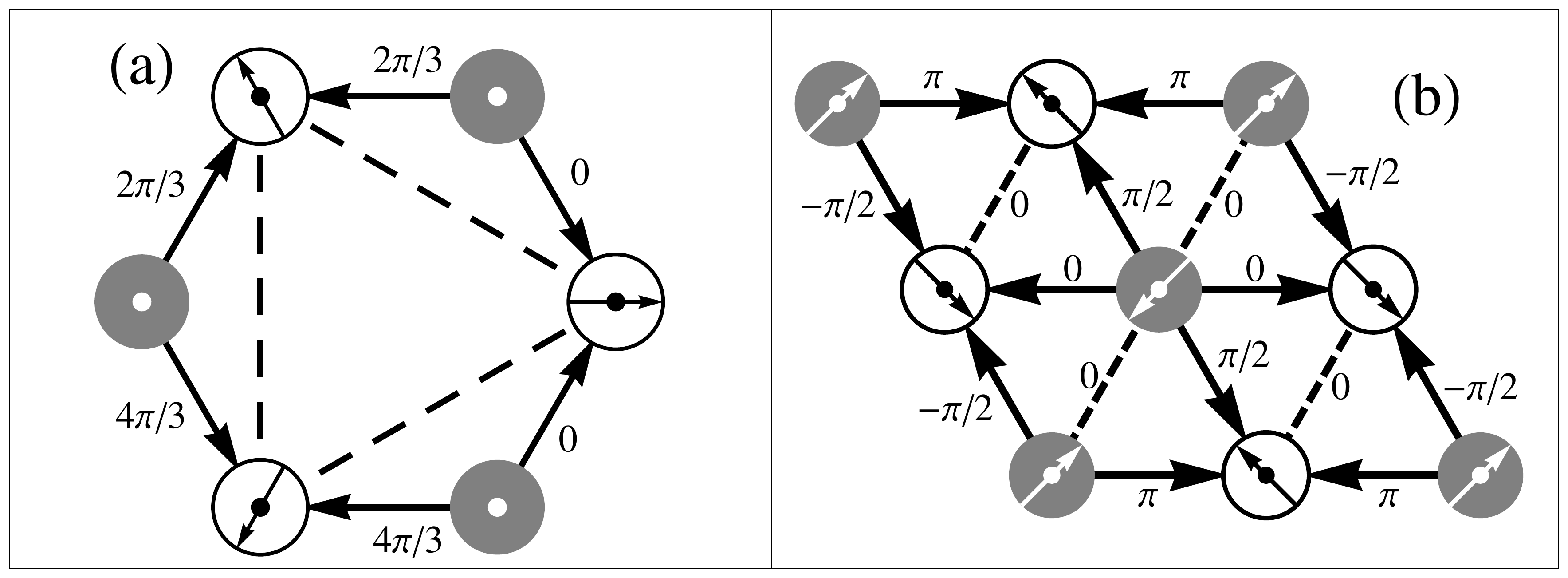}
\caption{\label{fig:hoppings} Peierls-like phases for hopping between sites (along the direction indicated by the black arrows on the bonds) in the tight-binding descriptions of the honeycomb (a) and triangular (b) lattice models. Arrows on the lattice sites show the phase $\phi_{\vec{R}}$ of the drive potential. Gray (white) circles denote A (B) sites as defined in the main text.
}
\label{fig2}
\end{figure}

\textit{Honeycomb Lattice---}
Our honeycomb lattice is formed from three red-detuned traveling waves with wavevectors $\vec{k}_1 =  k (0,1,0)$, $\vec{k}_2 =  -k/2 (\sqrt{3},1,0)$, $\vec{k}_3 =  k/2 (\sqrt{3},-1,0)$, shown in Fig.~\ref{fig:overview}.I(c).  The three beams interfere to produce a static lattice $V(\vec{r}) = V_\parallel(\vec{r})+V_\perp(\vec{r})$, with $V_\parallel(\vec{r}) = V_0\sum_{i<j}\cos\left(\vec{K}_{ij}\cdot\vec{r}\right)$ and  $V_\perp(\vec{r}) = -V_1 \sum_{i< j} \cos\left(\vec{K}_{ij}\cdot\vec{r}+\phis_{ij}\right)$, where $\vec{K}_{ij}\equiv\vec{k}_i-\vec{k}_j$, $\alpha_{ij}\equiv\alpha_i-\alpha_j$, and $V_{0,1}>0$. $V_\parallel$ and $V_\perp$ are lattices formed by, respectively, the $p$ and $s$ polarization components of the fields $\vec{E}_i$ (Eq. \ref{eq:Ei}); we fix the energy offset $\hbar \delta$ between A and B sites by using the polarization orientation $\theta$ to tune the relative intensities, and the ellipticity angles $\alpha_i$ to tune the relative displacements, of $V_\parallel$ and $V_\perp$ ~\cite{Luhmann2014}. 

The superlattice modulation is created by the interference of the $p$-polarized lattice beams with a circularly polarized coupling field at frequency $\omega+\Omega$ propagating normal to the plane: {\it i.e.} $\mathbf{E}= \hat{\vec{\sigma}}_- E_0 e^{i[kz- (\omega+\Omega)t]}$, where $\hat{\vec{\sigma}}_- = (\hat{\vec{x}} - i\hat{\vec{y}})/\sqrt{2}$.  The resulting potential
\begin{equation}
\label{eq:drivehoneycomb}
\Vmod=\Vmod_H \sum_{i=1}^{3} \cos\left(\vec{k}_i\cdot{\vec{r}+\Omega t+\gamma_i}\right)
\end{equation}
breaks time-reversal symmetry due to the winding of the phases $\gamma_i\equiv-\mathrm{Arg}[\hat{\vec{z}} \cdot (\hat{\vec{\sigma}}_- \times \vec{k}_i)]$ with wavevector $\vec{k}_i$.  $\tilde{V}$ modulates only the sites of the B sublattice, where the field $\sum_i \vec{E}_i$ of the static lattice has a $\hat{\vec{\sigma}}_-$ polarization component.  (The A sites instead have $\hat{\vec{\sigma}}_+$ polarization and thus, to lowest order, do not feel the time-varying drive potential.)  The B sites are all driven with the same amplitude, but with a phase that increases in steps of $2\pi/3$ on moving around a supercell containing three B sites, as indicated by the arrows on the white sites in Fig.~\ref{fig2}(a).

Using the prescription of Eq. \ref{eq:keff}, we find that the matrix elements for tunneling from a B to an A site acquire phases $0$, $2\pi/3$ and $4\pi/3$ (equal to the phase of the drive on the B site), illustrated on the solid links in Fig.~ \ref{fig2}(a). These tunneling phases lead to the fluxes shown in Fig.~\ref{fig:overview}.I(a), with a flux of $2\pi/3$ through B-A-B plaquettes. The model also has A-A couplings (not shown in Figs.~\ref{fig:overview}.I(a) or Figs.~\ref{fig2}(a)), but there is no flux through the corresponding A-B-A plaquettes.

\begin{figure}
\includegraphics[width=\columnwidth]{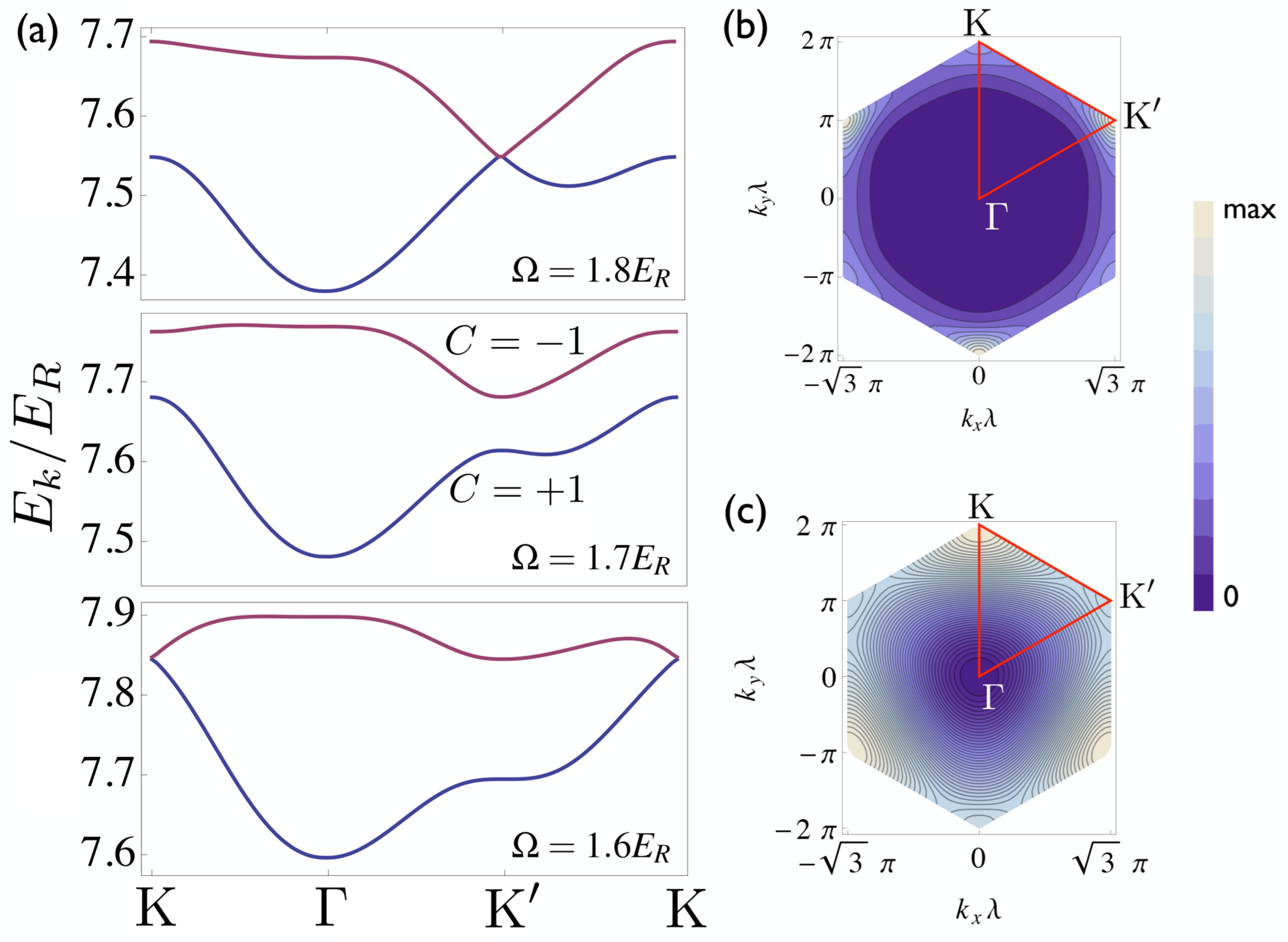}
\caption{(Color online.) (a) Floquet-Bloch bands for the honeycomb lattice with increasing modulation frequency $\Omega$.  The lattice parameters are (a) $V_0=10 E_R$, $V_1=0.5 E_R$ and $\tilde{V}_H=0.4 E_R$, where $E_R=\hbar^2 k^2/2m$.  Contour plots show the Berry curvature (b) and dispersion (c) of the ground band for $\Omega=1.71 E_R$.}
\label{fig:honeycomb}
\end{figure}

To determine the bandstructure of this dynamic superlattice, with time-periodic Hamiltonian $\hat{H}(t) = \hat{H}(t+2\pi/\Omega)$, we construct the Floquet states labeled by quasi-energy $\epsilon$~\cite{Shirley:1965dz}, with $-\hbar\Omega/2\leq \epsilon < \hbar\Omega/2$ \cite{SM}. The Floquet states are characterized by a conserved wavevector, $\mathbf{k}$, and give rise to two energy bands, there being two sites (A and B) in the magnetic unit cell.
When driven on resonance ($\Omega = \delta$ in the tight-binding picture), the dispersion relation features a single unsplit Dirac cone and mass gap at the other Dirac point. The unsplit Dirac point can be made to acquire a mass gap by tuning $\Omega$ away from $\delta$. Depending on the sign of this detuning, one obtains a model with Berry curvature of either opposite or equal signs at the two Dirac cones. 
When the bands are split such that the Berry curvature has the same sign in the lowest band at both Dirac cones, the bands are of non-trivial topology and have Chern number $|C| = 1$. For the example shown in Fig.~\ref{fig:honeycomb}(a), the bands are topological when $1.6 E_R<\Omega<1.8 E_R$.

\textit{Triangular Lattice.---}  To obtain topological bands that are better separated in energy, we construct a triangular lattice with uniform flux.  The laser configuration is illustrated in Fig.~\ref{fig:overview}.II(c).  A minimal set-up for the static lattice involves two retro-reflected beams with wavevectors $\vec{k}_{\pm 1} =  \pm k (0,1,0), \vec{k}_{\pm 2} =  \mp k/2 (\sqrt{3},1,0)$  [red arrows], which produce a potential
\begin{equation}
\label{eq:triangular}
V =- V_0\sum_{i=1}^2\cos^2(\vec{k}_i\cdot\vec{r}) + V_1 \cos(\vec{k}_1\cdot \vec{r})\cos(\vec{k}_2\cdot\vec{r}),
\end{equation}
where $V_0>0$.  The A-B offset $\delta$ is controlled by $V_1$, which is tuned to a value $\abs{V_1}\ll\abs{V_0}$ by setting polarizations to $\alpha_1=0$, $\alpha_2=\pi/2$, and $\gamma \ll 1$. To obtain the approximately six-fold symmetric lattice geometry shown in Fig. \ref{fig:overview}.II(b), an optional third lattice beam (dotted green arrows) can be included, adding a term $-V_0 \cos^2(\vec{k}_3\cdot\vec{r})$ to the static lattice $V$~\cite{footnote3}. Interference of the fields $\vec{E}_{\pm 1,\pm 2}$ with a linearly polarized beam at frequency $\omega + \Omega$ perpendicular to the plane generates a modulation
\begin{equation}
\label{eq:drivetriangular}
\Vmod=\frac{\Vmod_T}{\sqrt{2}} \left[ \cos(\vec{k}_1\cdot\vec{r})\cos(\Omega t) + \cos(\vec{k}_2\cdot\vec{r})\cos\left(\Omega t + \frac{\pi}{2}\right) \right].
\end{equation}
Here, the ($90^\circ$) phase lag between the two terms breaks time-reversal symmetry.  It is controlled by the position of one retro-reflecting mirror, which sets $\varphi_{-2}=\pi/2$ (relative to $\varphi_{+1,-1,+2}=0)$.

On sites of the triangular lattice [Eq. \eqref{eq:triangular}], located at $\mathbf{R}=m_1 \mathbf{R}_1 +m_2 \mathbf{R}_2$ with $\mathbf{R}_1=a (1,0)$ and $\mathbf{R}_2=a/2(1,\sqrt{3})$ ($a\equiv \lambda/\sqrt{3}$), the drive potential evaluates to
\begin{eqnarray}
\tilde{V}(\mathbf{R},t) =\frac{\Vmod_T}{2 \sqrt{2}}  \left[ (-1)^{m_2}+i (-1)^{m_1+m_2}\right] e^{i\Omega t}+{\rm c.c.}
\end{eqnarray}
Here, on A (B) sites, the amplitude and phase of the drive potential is proportional to $\pm 1 \pm i$ respectively, depicted as vectors in Fig.~\ref{fig:hoppings}. This drive gives rise to the tunneling phases $0$, $\pi/2$, $\pi$ and $3 \pi/2$ when hopping from B to A sites, according to Eq. \ref{eq:keff}.

\begin{figure}
\includegraphics[width=\columnwidth]{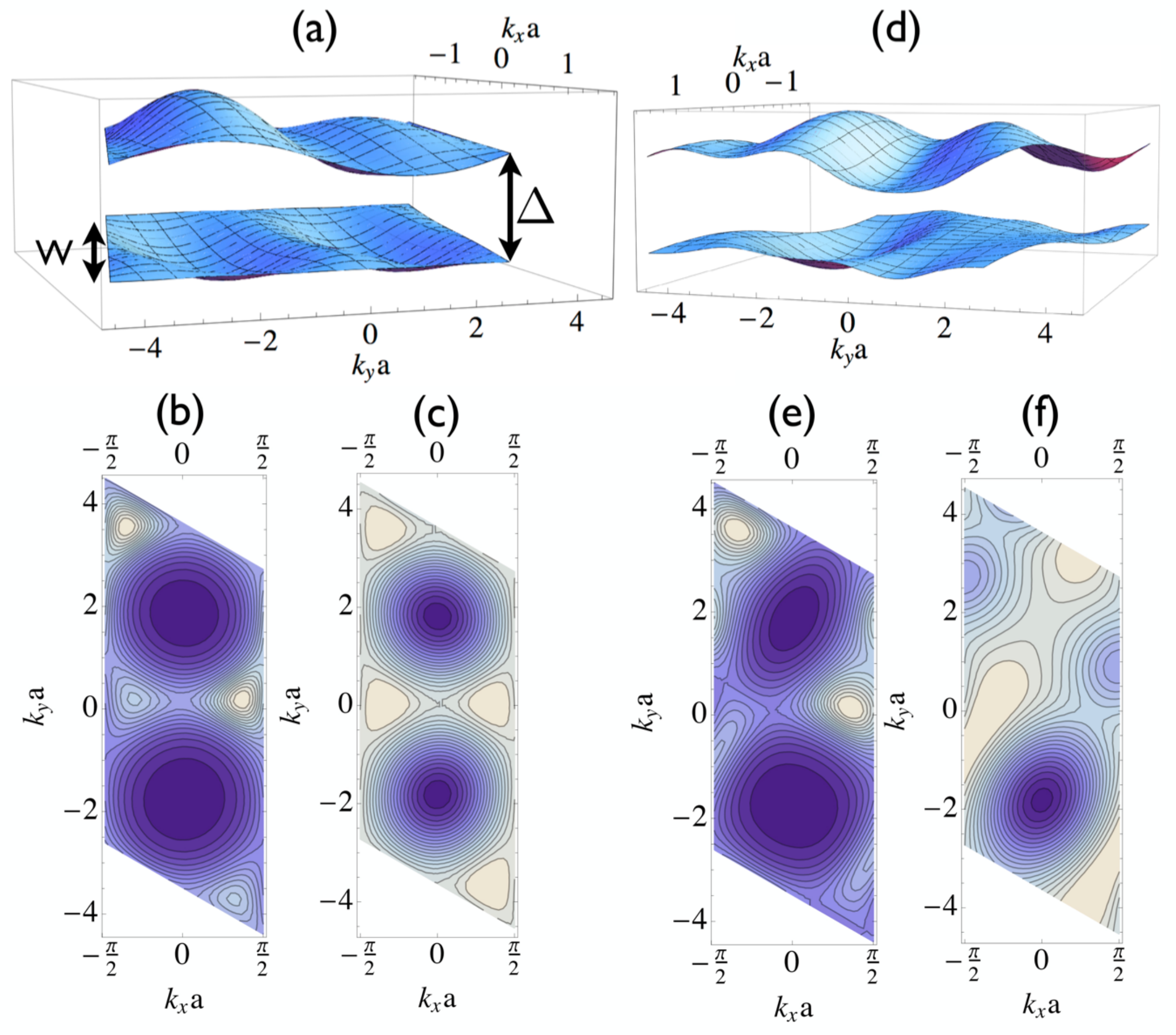}
\caption{(Color online.) Floquet-Bloch bands (top)  and the corresponding Berry curvature and dispersion relation of the lower band (bottom) for the triangular lattice with $\Phi=\pi/2$. In (d-f) the optional pair of laser beams [green dashed line in Fig.~\ref{fig:overview} II (c)] has been left out, whereas the lattice shown in (a-c) is approximately six-fold symmetric. The lattice parameters $(V_0,V_1,\hbar\Omega)$ are (a-c) $(6,0.6,0.85)E_R$ [(b-d) $(6,0.6,0.51) E_R$]. These result in a width of  the lowest band $W=0.0045 E_R$ [$W=0.02 E_R$] and a ratio of bandgap to bandwidth $\Delta/W \sim 1.9$ [$\Delta/W \sim 1.3$].}
\label{fig:triangular}
\end{figure}

The solution of the Floquet states shows a bandstructure very similar to that of the (time-independent) triangular lattice tight-binding model with $\pi/2$ flux per plaquette: the two bands are topological (with Chern numbers of $\pm 1$) and are separated by an energy gap $\Delta$ that is about twice the bandwidth $W$ [Fig. \ref{fig:triangular}(a)]. Small deviations of the bandstructure from an ideal isotropic triangular lattice with uniform flux are visible in Fig. \ref{fig:triangular}(a). These deviations come from a slight asymmetry between tunneling matrix elements for hopping from A to A  and B to B sites that depends on the precise shape of the superlattice potential.  Note that even without the third in-plane beam (dotted green lines in Fig. \ref{fig:overview}), one obtains well separated Chern bands [see Fig. \ref{fig:triangular}(b)].

\textit{Prospects.---}Signatures of the topological band structure could be detected using a variety of proposed techniques~\cite{Detection, PriceAbanin}.   E.g., as the bands are well separated at each point in momentum space for both dynamic superlattices, the Berry curvature could be fully characterized via the semiclassical dynamics of a wavepacket undergoing Bloch oscillations \cite{PriceAbanin}.

A consideration of the energy scales in the dynamic superlattices suggests that heating effects should be minimal. The bands have a
 natural energy scale set by the bare tunneling $J$ from A-to-A or
 B-to-B of the static lattice $V(\vec{r})$. For the honeycomb lattice, these next nearest neighbor tunnel couplings are small compared to the bare couplings
 $K$ from A-to-B in the absence of energy offset, $\delta =0$. Thus
 the ratio $|K_{\rm eff}|/J_{\rm eff}$  is widely tunable even in the regime of weak driving amplitude
 $\tilde{V}_{\vec {R}}/(\hbar\Omega) \ll 1$. For the triangular lattice, isotropic hopping amplitudes $|K_{\rm eff}| \simeq J_{\rm
   eff}$ are achieved for moderate driving amplitude $\tilde{V}_{\vec {R}}/(\hbar\Omega) \sim 1$, where multi-photon heating can remain weak~\cite{Hemmerich:2010ij}.
 The modulation frequency $\Omega \simeq E_{\rm R}$ is small compared to the gap to higher bands, so these bands do not contribute.  Furthermore, the absolute time-scales for tunneling are well within reach of current experiments. For example, for $^6$Li and a laser wavelength of $1~\mu$m, the bandwidth of the triangular lattice corresponds to a tunneling rate $1/\tau \sim W/h=130-590$Hz.
 
Our species-independent artificial gauge fields can benefit experiments with light alkali atoms, where alternative schemes involving Raman coupling of internal states cause rapid heating that precludes observing many-body physics~\cite{Cheuk:2012zr, Wei:2013kx}.  Both lithium and potassium offer bosonic and fermionic isotopes with fully tunable interactions. Exposing these species to a high magnetic flux may enable the study of novel strongly correlated states.

Notably, our scheme allows for almost \textit{adiabatic} loading into the lattice~\cite{Baur:2013ys}.  A protocol for preparing a Chern insulator could start by loading a gas of fermions into the lowest band of the static lattice. Subsequently, the coupling laser is ramped on but initially kept off resonance, at a drive frequency $\Omega_1\ll\delta$. Then, the drive frequency is swept to its final value $\Omega\approx\delta$. During this last step, the two coupled bands briefly touch at a single Dirac cone as they acquire a non-trivial topology. The gas can then be compressed to create a Chern insulator in the center of a trapped atomic cloud.

\acknowledgments{This work was supported by EPSRC Grant No. EP/I010580/1 and the A. von Humboldt Foundation.  We thank Ulrich Schneider and Immanuel Bloch for stimulating discussions.}

\newpage

\centerline{\textbf{Supplementary material}}
\vskip 5mm

\textit{Full Floquet calculation.}
We will now outline how one can calculate the Floquet-Bloch spectrum of dynamic optical superlattices and show that it gives results consistent with the tight-binding model description. For a time periodic Hamiltonian $\hat{H}(t)$, one can find a set of time-dependent quasi-stationary states
\begin{eqnarray}
\ket{\psi(t)}=e^{-i \epsilon t}\sum_j \ket{\psi_j} e^{- i j\Omega t}
\end{eqnarray}
with quasi-energy $\epsilon$~\cite{S:Shirley:1965dz}. These quasi-stationary states obey the eigenvalue equation
\begin{eqnarray}
\epsilon \ket{\psi_{j'}}= \sum_j \left(\hat{H}_{j-j'}-j \hbar \Omega \delta_{jj'}\right) \ket{\psi_j},
\end{eqnarray}
with
\begin{eqnarray}
\hat{H}_j=\frac{1}{T}\int_{0}^{T} dt\; e^{i j \Omega t} \hat{H}(t),
\end{eqnarray}

and we also define the Floquet Hamiltonian according to Ref.~\cite{S:Shirley:1965dz}
as
\begin{eqnarray}
(\hat{H}_F)_{jj'}=\hat{H}_{j-j'}-\hbar j \Omega \delta_{jj'}.
\end{eqnarray}
$T=2\pi/\omega$ is the oscillation period of the drive. In the cases we are considering in this paper, the single-particle Hamiltonians are of the general form
\begin{equation}
\hat{H}(t)=\frac{\hat{\mathbf{p}}^2}{2m}+V(\mathbf{r})+F(\mathbf{r}) e^{-i\Omega t}+F^*(\mathbf{r}) e^{i \Omega t},
\end{equation}
therefore we have $\hat{H}_0(\hat{\vec{p}})=\frac{\hat{\mathbf{p}}^2}{2m}+V(\mathbf{r})$, $\hat{H}_1=F(\mathbf{r})$, $\hat{H}_{-1}=F^*(\mathbf{r})$ and $\hat{H}_{|j|>1}=0$. The time independent potential $V(\mathbf{r})$ is invariant under lattice translations by direct lattice vectors $\mathbf{R}_1$, $\mathbf{R}_2$. For the time dependent potentials Eqs. \eqref{eq:drivehoneycomb}, \eqref{eq:drivetriangular} of the main text, $F(\mathbf{r})$, characterizing phase and amplitude of the time dependent drive, is a quasi-periodic function such that
\begin{eqnarray}
F(\mathbf{r}+\mathbf{R}_j)=e^{i \mathbf{G} \cdot \mathbf{R}_j} F(\mathbf{r}) \;\;\; \text{$j$=1,2}
\end{eqnarray} 
for some wave-vector $\mathbf{G}$. This quasi-periodicity enables us to perform a unitary (gauge) transformation in Floquet space in order to obtain a lattice periodic Floquet Hamiltonian (similar to what was done in \cite{S:Cooper:2011cr} for the flux lattice). For example with $(\hat{U})_{jj'}=e^{i j \mathbf{G}\cdot \mathbf{r}} \delta_{jj'}$, one obtains the lattice periodic Floquet Hamiltonian
\begin{eqnarray}
\hat{H}_F'=\hat{U} \hat{H}_F \hat{U}^{\dagger}.
\end{eqnarray}
The quasi-energies are then found by solving the eigenvalue problem
\begin{widetext}
\begin{eqnarray}
\left(
\begin{array}{ccccc}
\ddots & & & &\\
  & \hat{H}(\hat{\vec{p}}-\vec{G})-\hbar \Omega & e^{i \vec{G} \cdot \vec{r}} F^*(\vec{r}) & &\\
 & e^{-i \vec{G} \cdot \vec{r}} F(\vec{r})  & \hat{H}(\hat{\vec{p}}) &e^{i \vec{G} \cdot \vec{r}} F^*(\vec{r}) &\\
&  &e^{-i \vec{G} \cdot \vec{r}} F(\vec{r}) & \hat{H}(\hat{\vec{p}}+\vec{G})+\hbar \Omega & \\
& & & &\ddots
\end{array}
\right)
\left(
\begin{array}{c}
\vdots \\
\ket{\psi_{-1}}\\
\ket{\psi_0}\\
\ket{\psi_1}\\
\vdots
\end{array}
\right)=\epsilon \left(
\begin{array}{c}
\vdots \\
\ket{\psi_{-1}}\\
\ket{\psi_0}\\
\ket{\psi_1}\\
\vdots
\end{array}
\right).
\end{eqnarray}
\end{widetext}
Furthermore if an integer multiple $m$ of $\mathbf{G}$ is a reciprocal lattice vector
\begin{eqnarray}
m \mathbf{G}=n_1 \mathbf{K}_1+n_2 \mathbf{K}_2,
\end{eqnarray}
an alternative and, as it turns out, computationally more convenient choice for $\hat{U}$ is
\begin{eqnarray}
\hat{U}=e^{i \left(j \mod m\right) \mathbf{G}} \delta_{j j'}
\end{eqnarray}
In particular, for our proposed triangular lattice geometry, one has $m=2$ and for the honeycomb lattice $m=3$. After this gauge transformation, the Hamiltonian is lattice periodic and can be expanded in Floquet-Bloch wavefunctions. We then proceed to numerically calculate the band-structure of the lattice periodic Floquet Hamiltonian, by projecting into the subspace of the resonantly coupled pair of Bloch bands and keep as many Fourier modes as necessary to achieve convergence. Our results for the triangular lattice are shown in Fig. 4 of the main text. As can be seen in Fig. 4 (a) the three-beam triangular lattice it is possible to achieve a lowest band that resembles the lowest band of the corresponding tight-binding model with tunneling matrix elements of isotropic magnitude and flux $1/4$ per triangular plaquette. 
\begin{table}[htb]
\centering  
 \begin{tabular}{| l | l | l | l | l |}
    \hline
    Geometry & ${\cal A}_{\rm AA}/\cal{A}_{\rm AB}$ & ${\cal A}_{\rm BB}/\cal{A}_{\rm AB}$  & $\theta_{\rm BA}$ \\ \hline
    Triangular lattice & $\sqrt{2}$ & $\sqrt{2}$ & $\frac{\pi}{2} n$, $n=0,\ldots, 3$ \\ \hline
    Honeycomb lattice & $0$ & $\sqrt{3}$ &  $\frac{2\pi}{3} n$, $n=0,1, 2$\\ \hline
      \end{tabular}
          \caption{The first two columns show the amplitudes for hopping between A-A and B-B sites over the amplitude for A-B hopping for the driven triangular and honeycomb lattices described in the main text. Peierls-like phases for nearest-neighbor hoppings from B and A sites are also given (third column). Eqs. \eqref{eq:jeff}, \eqref{eq:keff} of the main text show how to relate ${\cal A}$ and $\theta$ to tunneling matrix elements. In Fig. \ref{fig2} of the main text it is shown how these Peierls-like phases are assigned to bonds.}\label{tbl:tunneling}
\end{table}

\begin{figure}
\includegraphics[width=\columnwidth]{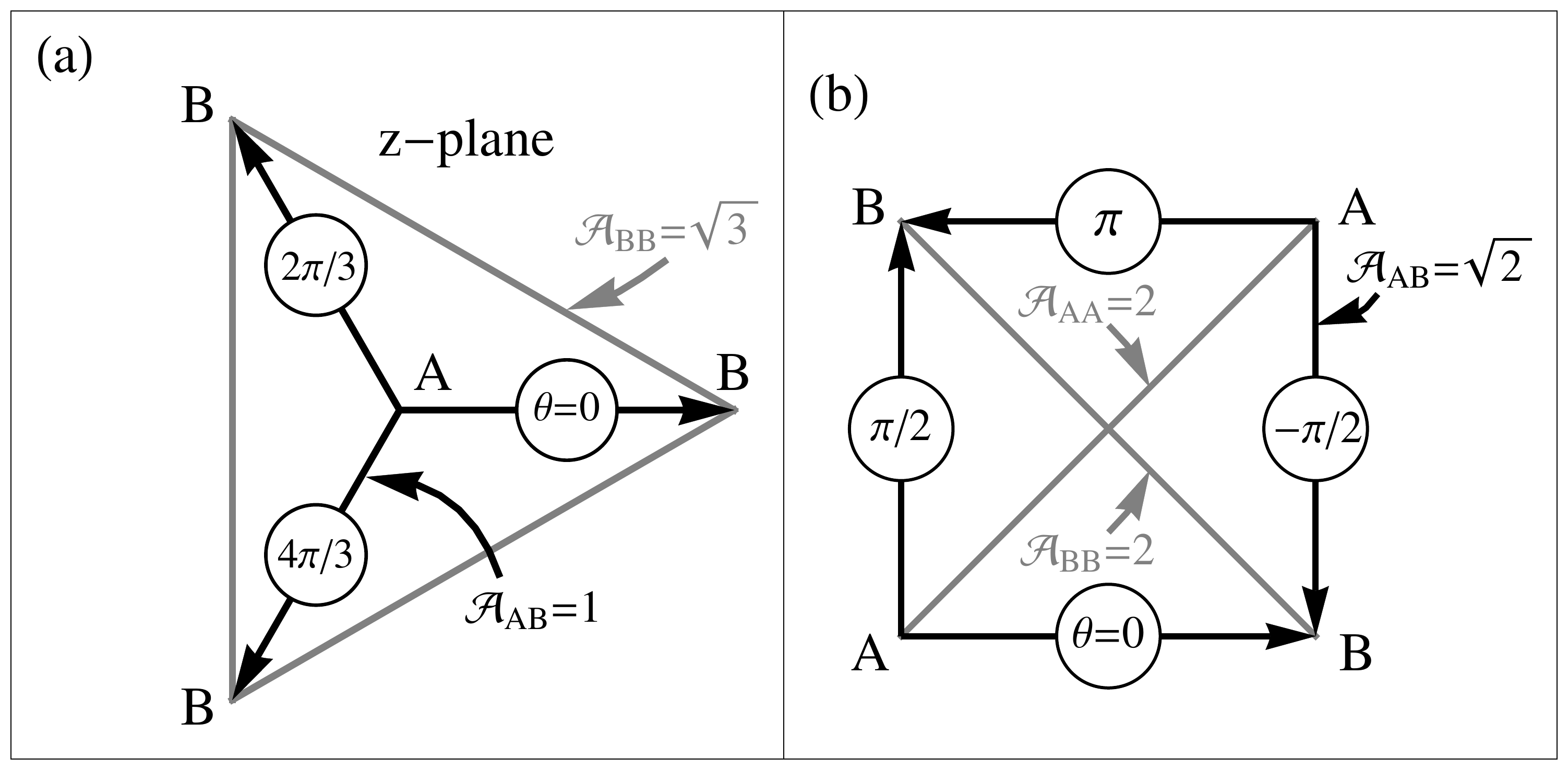}
\caption{Illustration of the derivation of the effective tunneling matrix element amplitudes (in units of $\tilde{V}_{\mathbf{R}}/\Omega$) and Peierls-like phases from the phase and amplitude of the drive potential on the A/B sites for both, honeycomb (a) and triangular lattice (b) geometries.}
\label{fig:hopping}
\end{figure}

\textit{Tight binding models.}
Here we describe the effective time independent tight-binding Hamiltonians discussed in the main text in more detail. To lowest order, the A-A tunneling matrix elements are not modified since drive amplitude on A sites vanishes. B sites are driven with amplitude $\tilde{V}_{\vec{R}}$ and with the phase pattern shown in Fig. \ref{fig:hoppings} of the main text. For the honeycomb lattice, the effective tunneling amplitudes are related to the bare matrix elements via
\begin{eqnarray}
J_{\rm eff}^{\rm AA} &\approx& J_{\rm AA}\\
J_{\rm eff}^{\rm BB} &\approx& J_{\rm BB} \times J_0(\sqrt{3} \tilde{V}_{\vec{R}}/\Omega)\\
K_{\rm eff}^{\rm BA} &\approx& K_{\rm BA} \times e^{i \theta_{\rm BA}} J_1(\tilde{V}_{\vec{R}}/\Omega).
\end{eqnarray}
Likewise, for the triangular lattice one finds the effective tunneling matrix elements (see Fig. \ref{fig:hopping})
\begin{eqnarray}
J_{\rm eff}^{\rm AA} &\approx& J_{\rm AA}\\
J_{\rm eff}^{\rm BB} &\approx& J_{\rm BB} \times J_0(2 \tilde{V}_{\vec{R}}/\Omega)\\
K_{\rm eff}^{\rm BA} &\approx& K_{\rm BA} \times e^{i \theta_{\rm BA}} J_1(\sqrt{2} \tilde{V}_{\vec{R}}/\Omega).
\end{eqnarray}


\begin{thebibliography}{10}
\bibitem{footnote3}
Interference with the other lattice beams can be avoided by slightly detuning the beam.
\end{thebibliography}
\end{document}